# Latent heat induced rotation limited aggregation in 2D ice nanocrystals

*Pantelis Bampoulis[1, 2, *], Martin H. Siekman[1], E. Stefan Kooij[1], Detlef Lohse[2], Harold J.W. Zandvliet[1], Bene Poelsema[1]*

[1] Physics of Interfaces and Nanomaterials, MESA[+] Institute for Nanotechnology, University of Twente,  P.O.Box 217, 7500AE Enschede, The Netherlands.

[2] Physics of Fluids and J.M. Burgers Centre for Fluid Mechanics, MESA[+] Institute for Nanotechnology, University of Twente,  P.O.Box 217, 7500AE Enschede, The Netherlands.

KEYWORDS: graphene, 2D ice, fractals, latent heat, molecular transport, heat transport, rotation limited aggregation

**The basic science responsible for the fascinating shapes of ice crystals and snowflakes is still not understood. Insufficient knowledge of the interaction potentials and the lack of relevant experimental access to the growth process are to blame for this failure. Here, we study the growth of fractal nanostructures in a two-dimensional (2D) system, intercalated between mica and graphene. Based on**

**our Scanning Tunneling Spectroscopy (STS) data we provide compelling evidence that these fractals are 2D ice. They grow while they are *in material* contact with the atmosphere at 20 °C and *without* significant *thermal* contact to the ambient. The growth is studied *in-situ*, in real time and space at the nanoscale. We find that the growing 2D ice nanocrystals assume a fractal shape, which is conventionally attributed to Diffusion Limited Aggregation (DLA). However, DLA requires a *low mass density* mother phase, in contrast to the actual currently present *high mass density* mother phase. Latent heat effects and consequent transport of heat and molecules are found to be key ingredients for understanding the evolution of the snow (ice) flakes. We conclude that not the local availability of water molecules (DLA), but rather them having the locally required orientation is the key factor for incorporation into the 2D ice nanocrystal. In combination with the transport of latent heat we attribute the evolution of fractal 2D ice nanocrystals to local temperature dependent Rotation Limited Aggregation (RLA). The ice growth occurs under extreme supersaturation, i.e. the conditions closely resemble the natural ones for the growth of complex 2D snow (ice) flakes and we consider our findings crucial for solving the "perennial" snow (ice) flake enigma.**

The multifarious morphology of snowflakes and their fascinating, highly symmetric appearances have attracted a lot of attention. Highly informative[1] and comprehensive[2] reviews have been written by Libbrecht. Today's understanding on the level of the crystalline *structure* of ice is well advanced and consensus has been reached on the importance of the hexagonal structure of natural ice, denoted as "ice-$I_h$", during initial growth[3,4,5]. However, the knowledge about growth *morphologies* and their evolution, i.e. on the growth *dynamics*, is still far from being understood. The snow crystal morphology diagram, showing crystal shapes as a function of temperature and supersaturation[1,2], is extremely rich. We focus on the high temperature - high humidity region, where fractal 2D platelets are found. The emerging morphology suggests signatures of diffusion limited aggregation (DLA)[6] or Mullins Sekerka[7] type growth instabilities. The role of heat transport due to temperature fluctuations caused by latent heat effects and that of molecular transport needed for incorporating molecules into the solid are still totally unknown. It has been suggested[1] that these transport issues play an important role in defining the ultimate morphology of the crystalline platelets.

Unfortunately, experimental access to latent heat effects and induced heat transport and their consequences on the emerging morphology of ice crystals at the molecular level is absent in 3D due to the lack of appropriate techniques. Even in 2D it is very hard to realize, especially under ambient conditions[2]. Xu *et al* have demonstrated in their pioneering work[8] that water can be aggregated between a mica substrate and a graphene sheet on top. Water intercalation between graphene (cover) and mica (substrate) have recently received tremendous attention[9-18]. Such intercalated water is in contact with the environment through defects in the graphene cover, which we will further refer to as B-defects[9]. At high relative humidity (RH) an about featureless image is obtained by using AFM (Atomic Force Microscopy). Upon a drop of the RH a brightness contrast develops: growing dark fractal shapes are surrounded by brighter areas. The dark areas represent a lower level and these depressions expand with time[9]. The height difference between these fractal depressions and their higher (brighter) environment amounts to about $0.37 \pm 0.02$ nm, which corresponds to the interlayer distance in hexagonal ice ($I_h$-ice)[8,10]. The shape of the depressions depends on the quality of the mica and the local thickness of the graphene cover sheet.

In this investigation we focus on the growth of the intercalated fractals reported first by Severin *et al*[9] and interpreted as de-wetting patterns. In contrast to ref[9], we identify these fractals as a 2D ice layer. The crystalline nature of the fractals is discussed in detail later. The surrounding higher level corresponds to a two bilayers thick water film[11]. In our experiment we study the growth of the 2D ice nanocrystals, under the conditions that lead to the growth of 2D snow (ice) flakes in nature. The intercalated ice crystals suggest growth of ice crystals under DLA conditions or Mullins-Sekerka type growth instabilities[6,7]. It is emphasized that DLA conventionally leads to low-density objects resulting from diffusing atoms present at a *low density* mass. When an atom hits the fractal aggregate it basically sticks and the aggregate expands. In contrast, in the current situation the aggregate grows from a very *high density* mass situation, the availability of the (in this case) molecules is not an issue. Rather, due to their anisotropic properties, the molecules stick only to the crystal provided they have the *required orientation*. We note that Mullins Sekerka type growth instabilities would lead to similar fractal appearances of the crystallites. They must be discarded too, based on similar density arguments. Further details are discussed below.

Exfoliated graphene (and Multilayer Graphene) was deposited onto a freshly cleaved muscovite mica (highest quality) at ambient conditions, by gently pressing a freshly cleaved HOPG (highly oriented pyrolytic graphite) flake (ZYA grade, MikroMasch) onto the mica. In our research, a strong preference for the highest HOPG quality was taken. Lower quality HOPG samples have a higher number of defects, resulting in higher density of fractals and interacting fractals. This can induce disturbances in their structure and make the analysis unnecessary complicated. The graphene application process has been also repeated inside a home built glovebox with controlled conditions and similar behavior has been observed. Extreme care has been taken to avoid contamination during the cleaving procedure by the adhesive tape[19]. Graphene was initially located using optical microscopy and the number of graphene layers was extracted from optical and atomic force microscopy (AFM) measurements[20]. For the optical investigation of the graphene flakes a DM2500 MH materials microscope (Leica, Germany) was used. AFM Imaging was performed in an environmental chamber using an Agilent 5100 atomic force microscope (Agilent) and HI'RES-C14/CR-AU probes (MikroMasch), with a nominal spring-constant of 5 N/m and resonance frequency of 160 kHz. The AFM scanner was operated inside an environmental cylindrical chamber. The RH was controlled by purging the environmental chamber with a controlled and adjustable $N_2$ flow or a wet air flow. The inlet of the N2 or wet air flow into the chamber was far away from the sample and AFM scanner in order to avoid disturbances during scanning induced by the flow. The wet air flow was realized by bubbling compressed air through distilled water. The humidity was controlled by adjusting the flow ratio. A RH of 0.1 – 1 % was realized using only the $N_2$ flow. The RH and Temperature of the environmental chamber were measured by a humidity sensor (SENSIRION EK-H4 SHTXX, Humidity Sensors, Eval Kit, SENSIRION, Switzerland) with an accuracy of 1.8 % between 10-90 % RH. The sensor was placed in a close proximity to the sample. The sample's plate temperature was controlled and measured using a Lakeshore 332 temperature Controller. Scanning tunneling microscopy (STM) and Spectroscopy (STS) was performed in air and in $N_2$ environment with an UHV variable temperature AFM/STM (BeetleTM, RHK Technology). The low humidity environment during the STM measurements was achieved by continuously purging the STM UHV chambers with $N_2$, the humidity sensor was placed in a close proximity to the sample. The water was purified to 18.2 MΩ·cm with a TOC of < 3ppb using a Mill-Q Reference A+ system (Merck Millipore). For the formation of fractal depressions a continuous $N_2$ flow (adjustable flow rate) was applied in the environmental. A 0.1 – 1 % RH is quickly reached and fractal depressions start to grow.

Before discussing our observations in greater detail we point out that the graphene/mica system has peculiarly advantageous properties and the presence of the graphene blanket, covering the water molecules on mica, is essential for several reasons. First, it is impermeable for water and thus prevents direct condensation and evaporation of water, from and into the ambient. Secondly, it also prevents the intercalated water to wet the probing AFM tip, which would give rise to strong perturbations of the water film. It therefore allows one to interrogate the *intrinsic* properties of the water/mica system making it accessible for experimental perusal. Last but not least, the intercalated water is confined between layered materials with strongly anisotropic properties. In both mica and graphene the thermal conductivity *along* the layers is much higher than that *perpendicular* to the layers. The in-plane and out of plane thermal conductivity of graphene at room temperature are 2000-4000 W/Km and 6 W/Km, respectively[21]. The values for muscovite mica are, respectively, 4.05 W/Km and 0.46 W/Km for in-plane and out of plane[22]. As a result the intercalated water is thermally "isolated" from the environment at 20° C. Local temperature fluctuations occurring in crystal growth/melting due to the evolution of the local heat of condensation/evaporation, which were suggested as being possibly crucial in the still poorly understood growth of ice/snow crystals[1], actually become accessible for the first time. These facts render the selected system predestined to study the yet uncharted consequences of temperature fluctuations at the molecular scale induced by latent heat effects for the growth of snow (ice) flakes. Below we find that these latent heat effects provide the missing link for understanding the formation of fractal ice nanocrystals. By accident, one obtains for free the easily accessible time-scale (minutes/ hours) needed to follow the molecular scale growth of 2D ice. Finally, we stress that the growth of the ice intercalated between graphene and mica is highly relevant for the natural growth of 2D snow (ice) flakes under the conditions indicated above. The graphene is only slightly hydrophobic and it acts as a "neutral" confinement. The mica is weakly hydrophilic. Due to the (almost) perfect lattice match between ice and mica (supplementary information[23]) the 2D ice condenses in an epitaxial fashion enhancing the visibility of the ice fractals. This 2D ice is substantially different than an ordinary bilayer of $I_h$-ice (typically grown on metal surfaces) and can be stable up to 300K[24].

Figure 1 shows AFM images of the graphene-water/ice-mica system, after a prolonged stay at a high RH (**A**), revealing the evolved fractals at 110 minutes after an initial drop of the RH (**B**). The darker fractals correspond to depressions with a height of 0.34-0.39 nm below the surroundings and they represent a 2D ice, while the unstructured, brighter areas correspond to a two bilayers thick water film. We also measured AFM phase images which demonstrate that in all cases the outermost graphene layer is imaged[8-10]. Indeed, the graphene sheet visualizes the water film underneath[8]. Figure 1**C** shows a typical STM image of a part of an evolved fractal. An overall hexagonal symmetry is observed. Figure 1**D** shows differential conductivities recorded on graphene above the 2D ice (blue curve) and the surrounding double water bilayer (red curve) of figure 1**C**. The differential conductivity of graphene above the double bilayer reveals a slight shift of the Dirac point towards the negative side (-50 mV), in accordance with density functional theory (DFT) calculations on graphene on mica[25]. The double water bilayer has effectively no net dipole moment and consequently zero effect on the electronic properties of the graphene cover[26]. In contrast the differential conductivity recorded on the graphene above the 2D ice crystal, reveals a large shift towards the positive side (+370mV). The observed spectra is a result of the unconventional character of the 2D ice crystal on mica[24]. The dipole orientation of the first 2D ice layer faces the mica surface[4,24,27], as a result, a net dipole moment is formed. Therefore, the graphene cover is doped in accordance to ref[25] and the Dirac point is shifted to the positive side. From the fact that these observations are fully in line with the DFT calculations in ref.[24], we conclude that our findings demonstrate that the fractals consist of 2D ice films on mica. A cartoon of the system is depicted in figure 1**E**.

Figure 2**A** shows sketches of the growing fractal at various times elapsed after the RH started to decay (for the AFM movie see supplementary information[23]). It demonstrates that mostly the extremities of the "younger" branches grow. Initially, the integrated area of the "trees" grows at a fast rate, which slows down continuously to finally get rather incremental, Figure 2**B**. A similar temporal evolution has been reported before by Severin et al.[9]. The fractal shape seems indicative of DLA[6,28], as a result of low mobility of molecules around the edges of the 2D ice crystal. Mullins Sekerka type growth instabilities would lead to a similar result[7].

Figure 3 shows a selected area with water films, intercalated between mica and, respectively, a single layer (top left) and a trilayer (bottom right) graphene sheet after prolonged stays at subsequently RH's of 50%, 1%, 11%, 17%, 25% and 50%, shown in panels **A** – **F**. A decay of the RH leads to the evolution of fractals (conform Figures 1 and 2). The ice fractals, intercalated between mica and a single, respectively, a trilayer of graphene, appear somewhat coarser in the latter case. Upon increasing the RH (**C** - **F**) the fractals assume initially a coarser and smoother shape and eventually their extremities retract (for the AFM movie see supplementary information[23]). Finally, the fractals disappear completely at RH = 50% leaving behind a double bilayer of intercalated water. A comparison of figures 3**A** and 3**F** demonstrates that the initial situation is completely restored after the cyclic RH variation has been completed. Figures 1, 2 and 3 demonstrate that the contact between the intercalated water and the environment is essential for the formation of fractal "trees" and the moderating B-defects define the location of their "trunks".

Figure 4**A** shows the growth of a fractal, following a drop of RH after a prolonged dwell at high RH. The fractals originate from the <u>same</u> defect and the first and second experiments are highlighted with blue and green, respectively. The overlapping sections are denoted by dark green. The subsequently grown fractals show only partial overlap, which proves that the branching is not dominated by defects, but rather is an *intrinsic property* of the growth of a 2D epitaxial ice crystal intercalated between mica and graphene. Further, evidence for the crystalline character of the fractals is provided in figure 4**B** representing the probability distribution of the branching angle. It clearly shows a predominant peak at 60°, characteristic of the hexagonal 2D ice. The emergence of a *crystalline* form of water is further corroborated by the fact

that evaporation from an intercalated liquid interface would lead to a spherical expansion of the depressed areas[12] and not to the observed fractal-shaped one. Due to its confinement between graphene and mica the surrounding double bilayer water film is "ice-like"[8], i.e the perpendicular ordering is high and long range lateral ordering is absent. It has been shown to be easily pushed away using the contact AFM mode[8] and STM [11], (see also supplementary information[23]).

The fractals are formed upon the evaporation of water into the ambient following a drop of the RH. Evaporation takes place exclusively at a location where a B-defect provides contact of the intercalated water with the atmosphere. The latent heat of vaporization of (bulk) water amounts to 2501 kJ/kg at 0 °C. The corresponding latent heat of freezing is 334 kJ/kg and thus the heat extracted from the system by evaporation of one molecule of water suffices to accommodate not less than seven water molecules in ice. For completeness we note that the different heat capacities (4.22 and 1.92 kJ/kg per K for water and ice) play only a subordinate role. The evaporation of a water molecule initiates a cold spot at the location of the facilitating B-defect. Since the intercalated water is thermally isolated from the layered surroundings (mica and graphene) the cold is effectively transported *parallel* to the water film (note, that cold loss due to radiation can be neglected, a crude estimate based on the Stefan-Boltzmann law results that the ice crystal will be within 5% of the environmental temperature after approximately 20 days). With the thermal conductivity of 3D ice being a factor of 4 higher than that of 3D water (2.2 W/Km versus 0.55 W/Km), the cold is transported predominantly through the ice. The close packed directions in the ice-crystal probably contribute most prominently. When the cold reaches the extremities of the ice crystal a part of it is also transferred to the surrounding water molecules, which leads to local cooling of the liquid. On these locations an ice molecule has more neighboring water molecules and a lower Kapitza resistance[29]. The Kapitza effect would enhance the growth rate at the protrusions. Relevant circumstantial evidence for this cooling is given in supplementary information[23]. The exchange of cold is most prominent at the extremities of the fractal, i.e. at those locations where the solid angle towards the liquid is relatively large and contact with more molecules in the liquid takes place. As a result the temperature gradient is high and thermal transport is facilitated. Also at these locations water molecules can more easily find a suitable lattice position to fuse with the ice and "annihilate" another part of the cold. In view of the high density of the water molecules and even though local clustering may limit the available molecules that can incorporate in the ice crystal, the finding of a suitable lattice position seems quite straightforward. However, the molecule also has to assume the required *orientation* to become promoted to the ice platelet. Such an orientation selection is much more restrictive and sufficient rotational freedom is key for becoming part of the 2D ice nanocrystal. Note, that the rotational freedom of the water molecules is constrained in an ice-like configuration and in this case is further hindered by the confinement[24,30]. Obviously rotation requires thermal activation and therefore the local "temperature" plays an important part. Especially around the older extremities more cold has been transferred into the surrounding liquid and the local temperature will thus be lower compared to an identical location around a younger extremity. Therefore, the orientation selection will be faster in the latter positions and growth will be faster at larger radial distances from the origin. In the DLA situation the growth of the latter extremities is faster due to shadow effects[28], active in the low mass density mother phase. In the current high mass density case and at a closer distance to the origin, the lower local "temperature" gives rise to more severe rotation restriction and thus lower local growth rates. Both situations lead to fractal shapes of the aggregates, but it appears that for the latter case RLA (Rotation Limited Aggregation) is a more appropriate acronym than DLA. Further evidence is provided in figure 5 and discussed later.

The growth of ice requires the removal of the second bilayer of water above the ice. Each incorporation of a water molecule into the 2D ice leads to the detachment of another molecule from the liquid double bilayer. Since the dipole orientation of the first ice layer is opposite to that of the liquid[27], an adsorbed water molecule (ad-molecule) has a weaker binding to the support. The second water bilayer destabilizes at the borders of the growing epitaxial 2D ice crystal. The location of the initial detachment from the double water bilayer is that where the crystal expands, i.e. at the very tip of the younger protrusions. The resulting ad-molecule can easily move across the ice by following, for steric effects, the C-C bonds underneath the graphene sheet[18] towards the B-defect and evaporate from there. Figure 4**C** sketches the above scenario.

For the reverse process occurring at excess RH, the fractal areas decrease: water molecules condense at the B-defect and induce a local hot spot, the heat will be transported preferably to the younger extremities, where ice molecules detach from the ice at the sharpest points, "annihilating" the heat. The local melting is further enhanced by the incorporation into the water bilayer of the ad-molecules that initially arrived from the ambient. This process causes a decrease of the local curvature and coarsening of the fractals. The scenario developed here accounts for all experimental observations presented above.

The growth instability gives rise to fractals, and the low edge mobility avoids quenching this shape instability[28,31,32]. Typically fractals grown at a higher temperature or at a lower rate have a coarser appearance[28]. Figure 5 shows images with fractals grown upon a drop of the relative humidity. The latter is achieved by introducing a flow of dry nitrogen through the box containing the sample. Figure 5 **A**, **B** and **C** show results for a nitrogen flow of, respectively, 15 L/min. 5 L/min and 1 L/min. The highest flow leads to thinner branches, while the flow reduction leads to coarser fractals. The lower flow leads to a slower evaporation of water molecules from the double bilayer and thus a lower growth rate of the fractals. The water molecules from the adjacent double water bilayer, that incorporate to the ice crystal, have sufficient time to find an energetically favorable orientation at a suitable site and the fractals become coarser.

Figure 5**D** shows fractals under monolayer graphene that have been formed at room temperature after exposure to 1 % RH. The same fractals acquire a coarser and smoother shape upon heating (the sample's plate temperature was increased to 100 ºC for 2 hours) and at constant RH (1 %), figure 5**E**. Figure 5**F** shows the overlay of (**A**) and (**B**). The white colour represents the increase of the fractals' width after exposure to higher temperature. A further expansion of the extremities is observed in agreement with Song et al[10]. The fractal expansion and coarsening is a result of exposure to a higher temperature.

In order to rationalize the observations it is noted that the situation is quite complex. The temperature effect is relevant for the water intercalated between mica and the graphene coversheet. Since the equilibrium amount of intercalated water decreases with increasing temperature the fractional area of 2D ice has a tendency to increase and vice versa the fractional area of the double bilayer thick water film will decrease. The time to reach equilibrium by equalizing the mass exchange through the B-defect will be very long due to its very small size. Note that in none of our experiments this situation is really reached. So growth of the fractals is expected upon temperature rise. At the same time the mutual exchange rate of molecules between the areas with one intercalated 2D ice layer, respectively a double bilayer of water is increased. However, both the rotation of the molecules to reach a suitable orientation for promotion into ice and the edge mobility necessary to find a more favorable site increase. As a result the fractal becomes coarser and smoother. This is further enhanced due to the increase of the ice-crystal temperature. The ice-crystal may undergo edge melting[a] at its extremities, which leads to enhanced edge mobility.

In summary, the processes leading to expansion and shrinkage of the 2D ice in contact with a double bilayer of water are complex as outlined above. The entire $H_2O$-system is intercalated between graphene and mica and constantly able to exchange material with the ambient at 20° C. The low thermal conductivity of both the bordering mica and graphene, normal to the interfaces, provides convenient *thermal* isolation of the intercalated water/ice from the ambient. This system enables one to observe in real time and real space the 2D molecular dynamics of the growth and melting of a 2D ice in contact with a double bilayer of water. Latent heat effects and their consequences for molecular motion, especially rotation, are found to be key ingredients to understand the appearance and disappearance of 2D ice crystals. Under the current prevailing high mass density circumstances the growth of fractals is not properly explained in terms of conventional diffusion limited aggregation (DLA). We rather attribute it to rotation limited aggregation (RLA) in combination with latent heat effects and subsequent heat transport. Our findings are of the highest relevance for the growth of planar snow (ice) flakes at relatively high temperature and high supersaturation, i.e. exactly those conditions that lead to fractal (dendritic) snow (ice) flakes (cf. morphology diagram[1,2]). Therefore the current findings form an essential ingredient for solving the longstanding enigma around the growth of complex natural snow (ice) flakes. We suggest that latent heat effects, may well have important

---

a The melting of 3D crystals often goes through a stage of surface (2D) melting[33,34]. In analogy one may presume that the 1D borders of 2D crystallites also melt before the 2D "bulk" does.

implications for crystal and thin film growth. This holds in particular for anisotropic materials and ultrathin films with a poor thermal contact to the substrate.

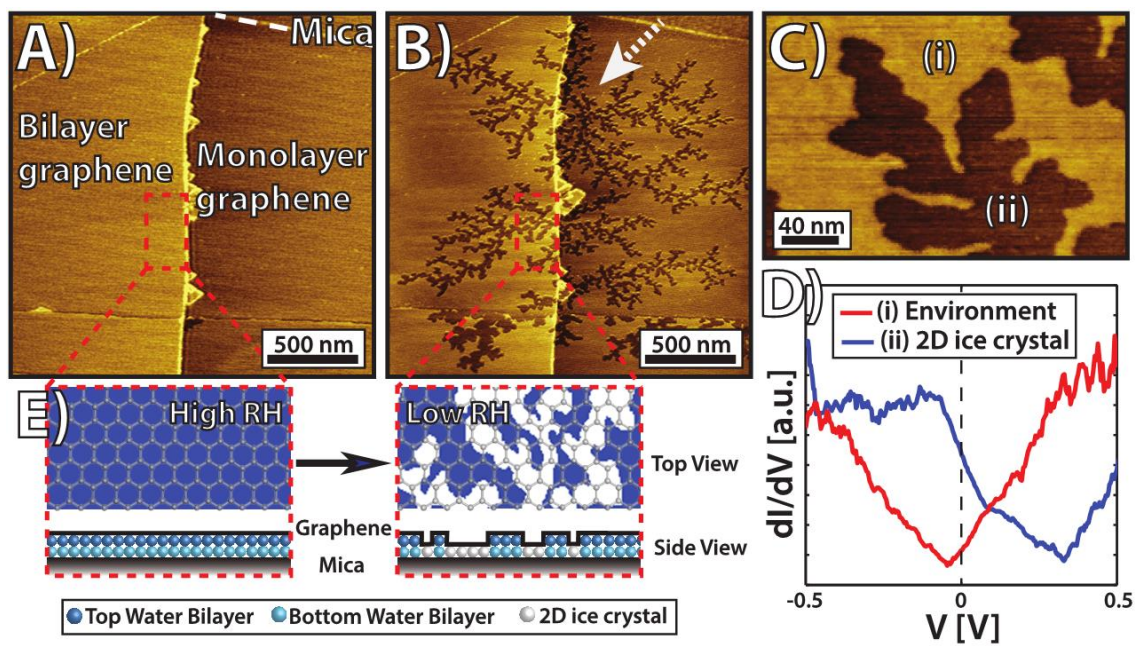

**Figure 1.** (**A**) and (**B**) AFM images of the surface of a graphene sheet on top of mica with an intercalated ultrathin water film. The thickness of the graphene varies locally (1 and 2 layers). (**A**) After long exposure to 50% RH (B) 110 minutes after the drop of the RH. It typically takes a few minutes to reach the new RH. (**C**) STM image of a branch of a fractal. The sample bias is 0.5 V and the tunnel current is 0.2 nA. (**D**) Differential conductivity recorded at the environment of the fractal, red curve (region (i) of figure (**D**)) and on the fractal, blue curve (region (ii) of figure (**D**)). The sample bias for both curves is 0.5 V and the tunnel current is 0.2 nA. (**E**) A schematic representation of the graphene/water/mica system (out of scale).

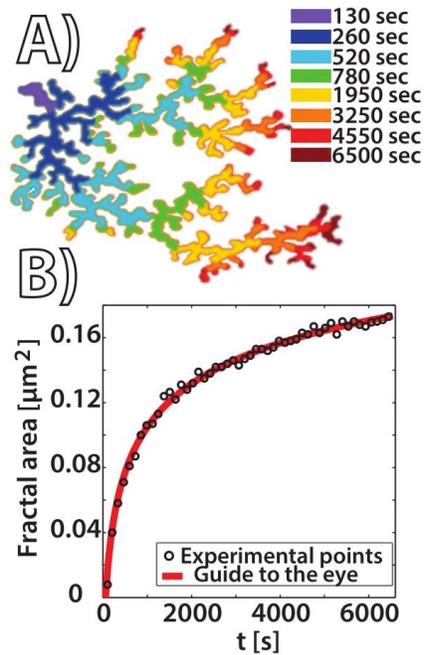

**Figure 2.** (**A**) The fractal indicated with the white arrow in figure 1B at different times after introduction of the RH decay. (**B**) The areal growth of the same fractal versus time after starting humidity decay. The time is counted from the start of the growth of the fractal after a purging-rate dependent incubation time of minutes.

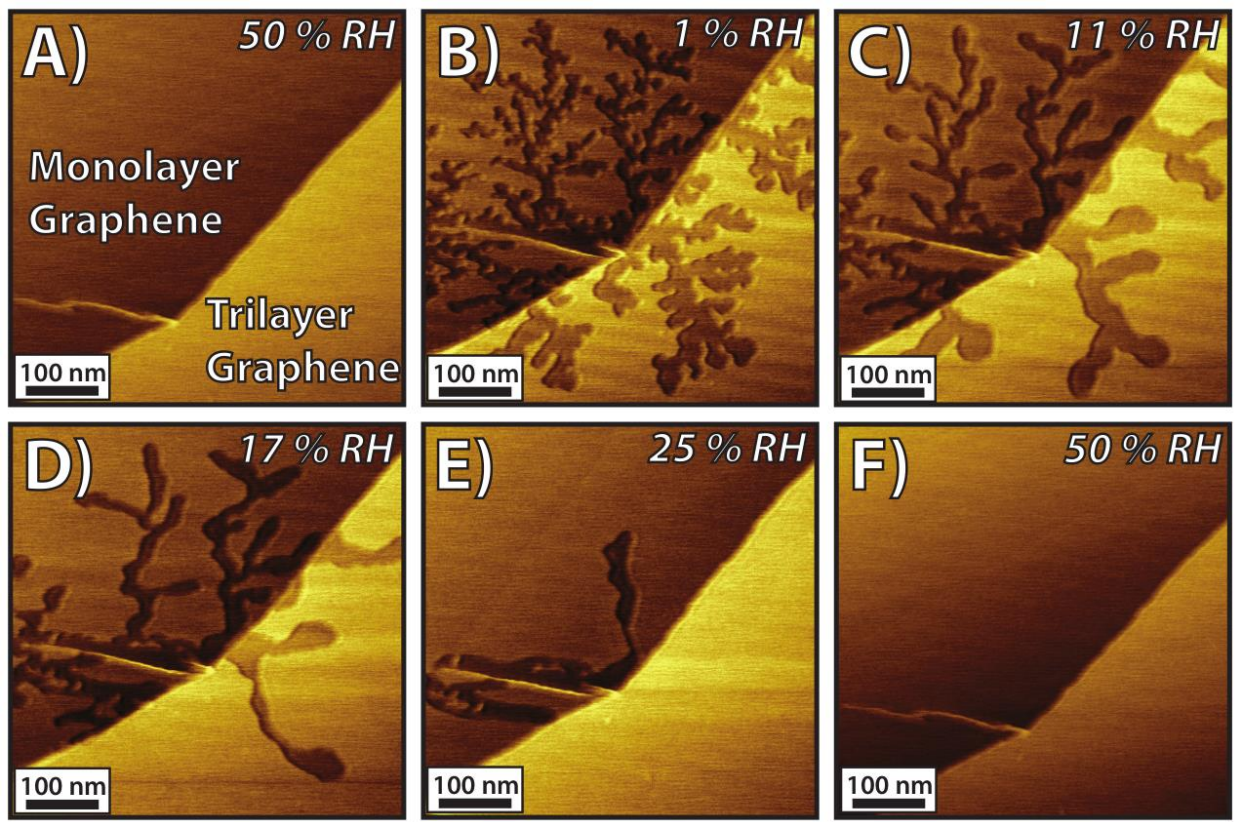

**Figure 3.** AFM images of graphene on top of mica with an ultrathin intercalated water film. The images contain sections with monolayer and trilayer thick graphene sheets. The images (**A**-**F**) were obtained subsequently at the indicated RH's. The bright narrow feature in all images is a graphene defect, probably a wrinkle, allowing contact with the ambient.

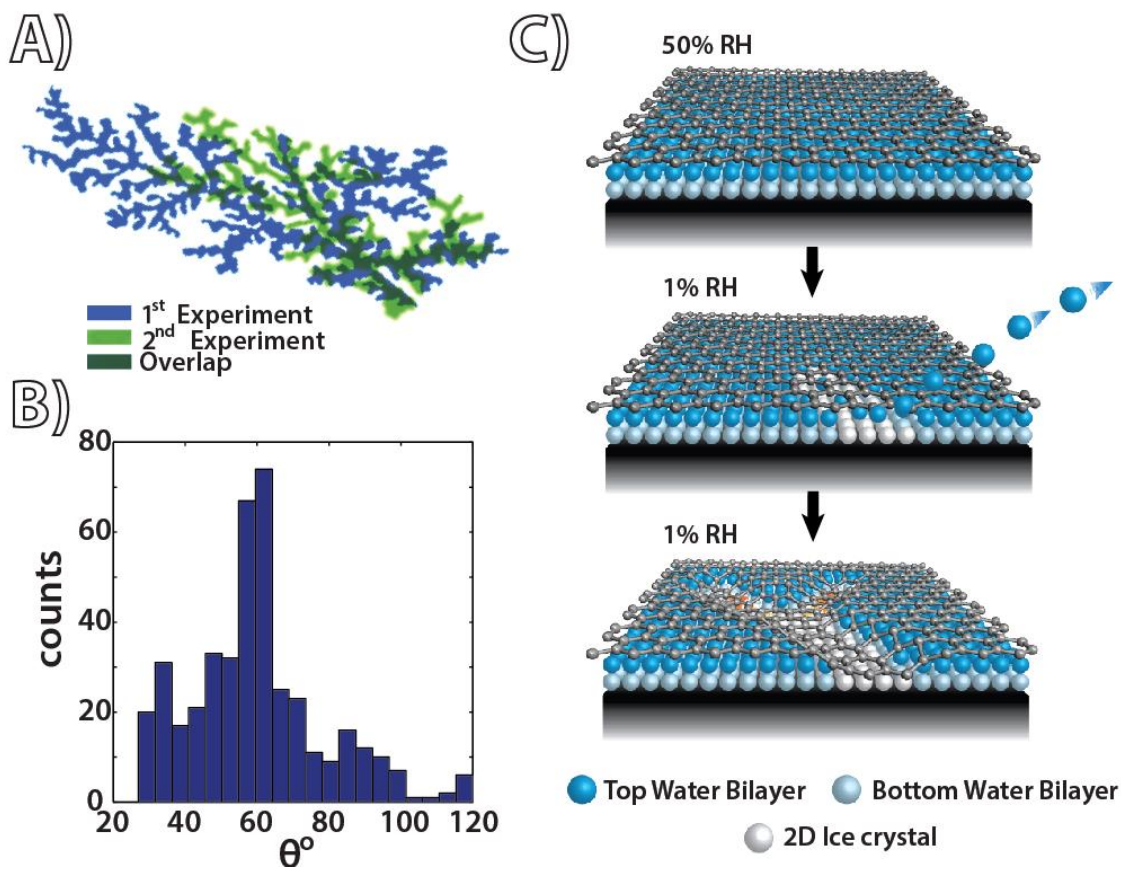

**Figure 4.** (**A**) Two fractals originating from the same defect. They are overlaid from two subsequent experiments following long dwells at, subsequently, high RH – low RH (blue area) - high RH – low RH (green area). The dark features represent overlapping areas. The fractals were formed after purging with $N_2$ at 5 L/min. (**B**) Probability distribution of the branching angles of the fractals. (**C**) A simple cartoon of the intercalated water film at different times (out of scale). It represents the double bilayer intercalated between graphene and mica at high RH. At a drop of the RH, water molecules from the double bilayer evaporate through the contact point to the environment. A 2D ice platelet nucleates and propagates through the bottom water bilayer. A 2D ice crystal is formed surrounded by a double water bilayer.

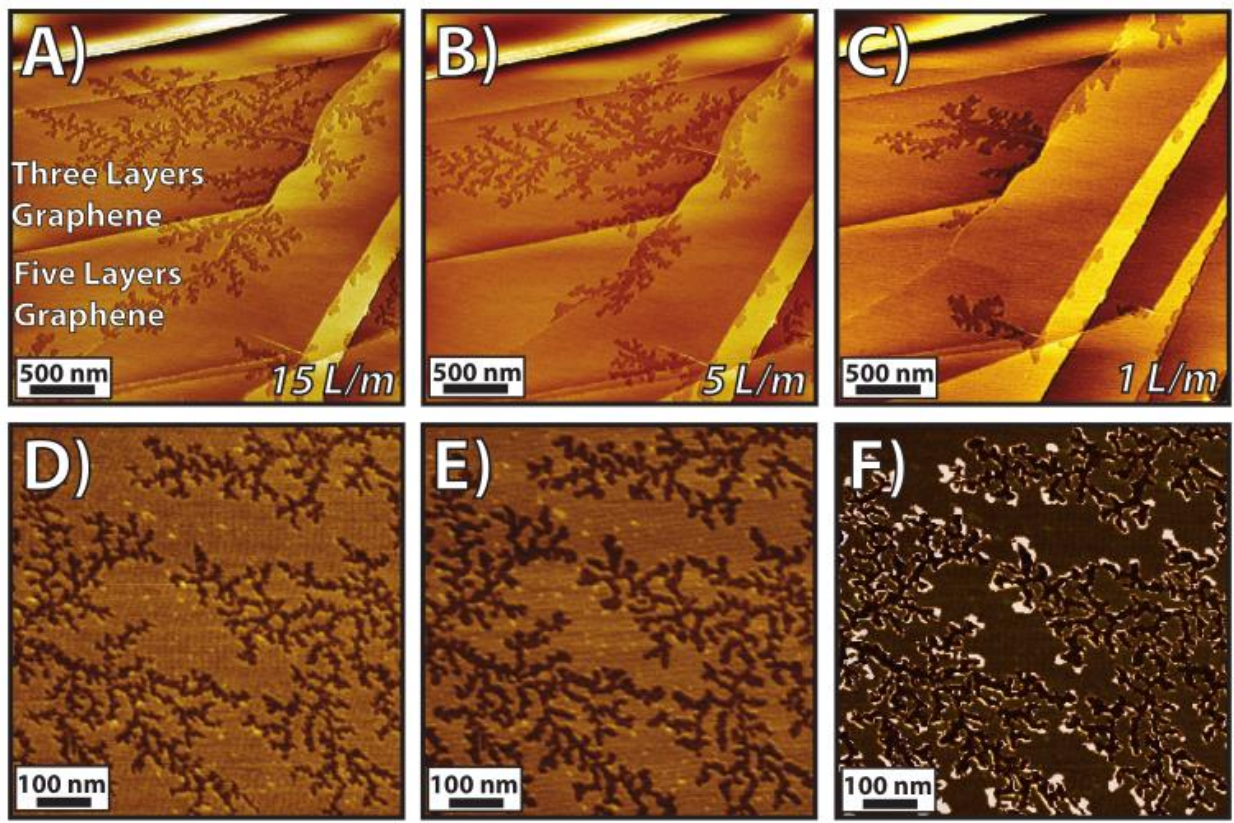

**Figure 5**. Fractals that emerged from the same B-step defect after prolonged stays at low RH induced by a decay of the RH, induced by N$_2$ purging at a rate of 15 L/min (**A**), 5 L/min (**B**) and 1 L/min (**C**). The initial situation, complete intercalated double bilayer water film, was identical for all cases. (**D**) Ice fractals under monolayer graphene at 20 ºC ambient temperature. (**E**) The same fractals after 1 hour exposure to higher temperature (the sample plate was heated to 100 ºC). The fractal branches appear coarser and somewhat smoother in the latter case. (**F**) Shows the overlay of (**D**) and (**E**), the white color indicates the increase of the fractal branches width after exposure to a higher temperature.

AUTHOR INFORMATION

**\*Corresponding Authors**

Pantelis Bampoulis (p.bampoulis@utwente.nl)

**Author Contributions**

Pantelis Bampoulis conceived the project, designed the experiments, analyzed the data and wrote the manuscript. Bene Poelsema oversaw the work, provided the framework for the interpretation and critical feedback. Martin H. Siekman helped with the experiments. E. Stefan Kooij, Detlef Lohse and Harold J.W. Zandvliet, provided feedback and helpful discussions. All authors have given approval to the final version of the manuscript.

**Funding Sources**

Dutch Organization for Research (NWO, STW 11431)

ACKNOWLEDGMENT

P. Bampoulis would like to thank the Dutch Organization for Research (NWO, STW 11431) for financial support.

**Supplementary Information 1. Epitaxial 2D-water on mica**

Due to a perfect epitaxial relationship between mica and (rotated) palladium (Pd) the latter nicely forms epitaxial crystallites on mica[S1]. The oxygen-oxygen distance in ice is equal to that of the Pd nearest neighbor distance with an accuracy of only a few promilles. Therefore, 2D ice strongly tends to form epitaxial films (ice) on mica as well. Obviously, the bond of Pd with mica is much stronger than that between water and mica. Mica is only weakly hydrophilic, but with a very small effective lattice mismatch the mica may still guide the azimuthal orientation of the ice. Therefore, 2D ice is expected to grow in an epitaxial fashion on mica in analogy with Pd/mica.

**Supplementary Information 2. Nanomanipulation of the Fractal's environment**

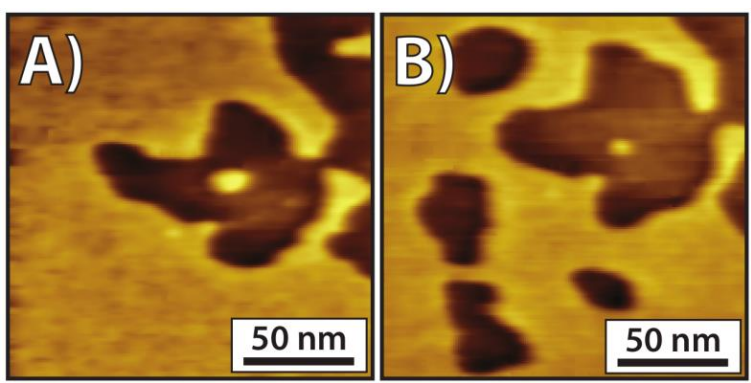

**SF1.** STM topographies before (A) and after (B) grid STS measurement. The set point was 0.5V and 0.2nA.

Figure SF1 shows topographs of a depression before (A) and after (B) doing grid STS. It is shown that after STS irregular depressions at the environment of the fractal are created and, similar to the fractals, are stable over time. This indicates that the tunnelling electrons can pass through the graphene cover and interact with the water underneath. Most probably these depressions are formed due to local heating induced by the STS measurement. However, due to the crystalline structure of the ice fractal (fractal depressions) and the epitaxial relation to the hydrophilic mica, we were unable to directly manipulate the ice layer by doing STS, in line with ref[8,11]. Though as one can see from our STM images before and after grid STS (figure SF1) the edges of the fractal have been changed. We have observed a similar effect in our variable temperature experiments (see figure 5), where the fractal becomes smoother and coarser when exposed to a higher temperature. In this case we attributed the coarsening of the fractal to an increase of both the rotational freedom of the molecules and edge mobility due to heating. Furthermore, due to the increase of the ice-crystal temperature, the ice-crystal undergoes edge melting at its extremities, which leads to enhanced edge mobility and rotational freedom. As stated before, this is in analogy to the surface 2D melting of 3D crystals[33,34], similarly the 1D borders of the ice crystals also melt before the 2D "bulk" does.

**Supplementary Information 3. Transfer of cold to the double bilayer of intercalated water**

Severin et al[9] have reported the growth of fractals in the double bilayer of water, intercalated between graphene and mica, upon purging their system with dry air after starting at ambient conditions. In Fig. 2D of their paper, they show an image obtained after purging the system for 14 hrs. Besides a fully developed fractal the image also shows a high density of much smaller fractals, which make no direct contact to the large fractal at all. We attribute these smaller fractals to nucleation and growth of smaller ice fractals. During the 14 hours of purging the water continues to evaporate at a slow pace due to slow diffusion of water molecules towards the "B-defects" and disappear into the ambient by evaporation. Therefore the transport of cold towards the extremities of the ice fractals continues and cold is transferred to the double bilayer of water. This leads to a continued lowering of the temperature of the double water-bilayer with as a consequence an ever growing supersaturation. At sufficiently high super-saturation, nucleation of 2D ice crystals takes place and subsequently the nuclei start to grow. The high density of nuclei reveals low diffusivity of the double bilayer water and, also in line with our observations, the homogeneous spatial distribution of the nuclei confirms that the observations are characteristic of the intrinsic properties of the system and no important influence of defects (if at all) is seen here. The latter is illustrated in Fig. 1F of[9]. From the density of the ice nuclei, as well as their size, as a function in increasing distance one easily distinguishes the consequences of a positive temperature gradient from right (low) to left (high).

**Supplementary References**

S1. S. Buchholz, H. Fuchs, and J. Rabe, Surface structure of thin metallic films on mica as seen by scanning tunneling microscopy, scanning electron microscopy, and low-energy electron diffraction. *J. Vac. Sci. Technol. B 9*, 857-861 (**1991**).